\documentclass[square, numbers]{article}

\usepackage[final]{neurips_2020}
\usepackage[utf8]{inputenc}
\usepackage[T1]{fontenc}
\usepackage{hyperref}
\usepackage{url}
\usepackage{amsmath, amssymb, amsfonts}
\usepackage{microtype}
\usepackage{graphicx}

\hypersetup{
  pdftitle={Provability vs. Execution: A Comment on “Consequences of Undecidability in Physics on the Theory of Everything”},
  pdfauthor={Evan Redden},
  colorlinks=true,
  linkcolor=blue,
  citecolor=blue,
  urlcolor=blue
}

\title{Provability vs. Execution: A Comment on “Consequences of Undecidability in Physics on the Theory of Everything”}

\author{
  Evan Redden \\
  ORCID: 0009-0008-6389-1871 \\
  Western Governors University \\
  \texttt{eredde5@wgu.edu}
}

\begin{document}

\maketitle

\begin{abstract}
Recent work by Faizal et al. (2025) claims that Gödelian undecidability of non-algorithmic truths in our universe imply the impossibility of a formal, algorithmic simulation of the universe. This paper clarifies the distinction between epistemic incompleteness: limits on what can be proven within a formal system, and ontological incompleteness: limits on what can exist or be computed by that system. Using Conway’s Game of Life as a Turing-complete example, I demonstrate that undecidability constrains provability but not computability or execution. Unless physical phenomena require the resolution of undecidable propositions, incompleteness alone does not imply a guaranteed failure in execution. Thus, the claim that the universe cannot be simulated lacks empirical and logical justification without evidence of hypercomputation in nature.
\end{abstract}

\section{Introduction and Background}
The pursuit of a unified, formal description of the universe has long motivated the development of both physical and computational theories of reality. Recent arguments have applied results from mathematical logic: specifically Gödel’s incompleteness theorems \cite{godel1931formal}, Tarski’s undefinability theorem \cite{tarski1936truth}, and Chaitin’s information-theoretic limits \cite{chaitin2004meta} to suggest that any formal, algorithmic description of the universe must be inherently incomplete. Faizal et al.~(2025) extrapolate this reasoning further, concluding that such incompleteness implies the impossibility of a formal simulation of the universe itself \cite{faizal2025undecidability}.

This paper challenges that conclusion by distinguishing between two fundamentally different limitations: \emph{epistemic incompleteness}, which concerns what can be known or proven within a formal system, and \emph{ontological incompleteness}, which concerns what can physically exist or be computed by that system. Although Faizal's results constrain the former, they do not necessarily impose limits on the latter. The distinction is critical: an incomplete theory may still describe a complete and executable physical process.

\section{Review and Conceptual Clarification}
Faizal et al.~construct a formal language, $\mathcal{L}_{QG}$, corresponding to a formal system $\mathcal{F}_{QG}$ that expresses statements about a prospective theory of quantum gravity. They correctly note that any sufficiently expressive formal system is incomplete: there exist true statements within $\mathcal{L}_{QG}$ that cannot be proven within $\mathcal{F}_{QG}$ itself. From this, they infer that a computational model of the universe based on $\mathcal{F}_{QG}$ cannot reproduce all physical truths, since some truths are undecidable within the system \cite{faizal2025undecidability}.

However, this reasoning conflates what can be \emph{proven} within a formal theory with what can be \emph{executed} by an algorithmic process. A formal system may be unable to prove all statements about its own behavior, yet the rules that generate that behavior can still be mechanically applied without contradiction. The process of computation does not require the system to resolve its own undecidable propositions in order to proceed to its next state.

An illustration of this is Conway’s Game of Life: a cellular automaton defined by a small set of deterministic rules \cite{conway1970game}. Despite being governed by a finite algorithm, the Game of Life exhibits Turing-complete behavior, meaning that undecidable questions (e.g., whether a configuration will ever stabilize) arise within it. These undecidable truths do not prevent the system from evolving; they merely restrict what an observer can predict or prove about its evolution.

\section{Hypercomputation and Ontological Limits}
To argue that our universe cannot be simulated algorithmically, one must demonstrate that physical processes require computational resources beyond those available to a Turing machine \cite{turing1937computable}. In other words, the dynamics of the universe must depend on the resolution of an undecidable problem in order to advance from one state to another. This would entail the existence of \emph{hypercomputation} in nature: physical processes capable of computing non-Turing functions or solving the Halting Problem \cite{copeland1998hypercomputation}.

Although several theoretical constructs have been proposed as candidates for hypercomputational systems, none have been shown to be physically realizable due to the Church-Turing thesis. Without empirical evidence of such mechanisms, the claim that undecidability forbids simulation remains speculative \cite{martin2006hypercomputation}. In the absence of hypercomputation, the undecidable propositions arising from formal descriptions of physical law reflect epistemic limits of knowledge, not ontological limits on what the universe can compute or produce.

\section{Discussion and Conclusion}
The application of Gödelian results to physics must be interpreted with care. These theorems delineate the boundaries of formal reasoning, not of physical causation. A consistent algorithmic theory of physics may indeed be incomplete, unable to justify itself or prove all truths within its own language; but it may nonetheless generate a fully consistent, evolving universe. Incompleteness limits what we can \emph{know} about the system, not what the system can \emph{do}. 

Unless the laws of nature can be shown to rely on hypercomputational processes, there is no logical or empirical basis to conclude that the universe cannot be simulated by a formal, algorithmic system.

\section*{Acknowledgements}
The author acknowledges the work of Faizal et al. (2025) \cite{faizal2025undecidability}, which motivated the present analysis. This paper was completed independently as a personal research project and was not affiliated with or supported by any formal research program at Western Governors University. The author also acknowledges the use of two large language models, ChatGPT and Gemini, for assistance with formatting the manuscript into a scholarly style and for refining specific illustrative language, such as the analogy using Conway’s Game of Life. The central thesis and core arguments of this paper were independently conceived and authored by Evan Redden. Any errors or misinterpretations remain the sole responsibility of the author.

\vspace{1em}

\setcitestyle{numbers}
\bibliographystyle{apalike}
\bibliography{references}

\end{document}